\documentclass{article}
\usepackage[preprint]{neurips_2025}

\usepackage[utf8]{inputenc} 
\usepackage[T1]{fontenc}    
\usepackage{hyperref}       
\usepackage{url}            
\usepackage{booktabs}       
\usepackage{amsfonts}       
\usepackage{nicefrac}       
\usepackage{microtype}      
\usepackage{xcolor,colortbl}
\usepackage{color}
\usepackage{soul} 
\usepackage{graphicx}
\usepackage{multirow}
\usepackage{xspace}
\usepackage{amsmath}
\usepackage{comment}
\usepackage{xcolor}
\usepackage{soul}
\usepackage{etoolbox} 
\usepackage{colortbl}
\usepackage{caption}
\usepackage{comment}

\newcommand{\SysName}{DeMT\xspace}

\title{Device-Guided Music Transfer}

\author{%
  Hung Manh Pham \\
  Singapore Management University\\
  \texttt{hm.pham.2023@phdcs.smu.edu.sg} \\
\And
Changshuo Hu\\
Singapore Management University\\
\texttt{cs.hu.2023@phdcs.smu.edu.sg}\\
\And
Ting Dang\\
University of Melbourne\\
\texttt{ting.dang@unimelb.edu.au}\\
\And
Dong Ma\\
Singapore Management University\\
\texttt{dongma@smu.edu.sg}\\
}

\begin{document}

\maketitle

\begin{abstract}
Device-guided music transfer adapts playback across unseen devices for users who lack them. Existing methods mainly focus on modifying the timbre, rhythm, harmony, or instrumentation to mimic genres or artists, overlooking the diverse hardware properties of the playback device (i.e., speaker). Therefore, we propose \SysName, which processes a speaker's frequency response curve as a line graph using a vision-language model to extract device embeddings. These embeddings then condition a hybrid transformer via feature-wise linear modulation. Fine-tuned on a self-collected dataset, \SysName enables effective speaker-style transfer and robust few-shot adaptation for unseen devices, supporting applications like device-style augmentation and quality enhancement. Audio samples are available at \url{https://anonymous.4open.science/w/music-examples/}. 
\end{abstract}

\section{Introduction}

Music is an essential part of human culture, shaping emotions, memories, and experiences. The way music is perceived, however, can be highly influenced by the playing device, as different devices introduce different spectral speaker characteristics. Accordingly, recent advancements in earphones and headphones have introduced a vast array of devices, spanning diverse hardware designs. This variety can largely impact music perception and playback quality even when the digital input file is the same. For music producers, audio researchers, and general users, such variations present challenges in assessing and optimizing music for different speakers that align with human ear perception. Therefore, speaker-guided music transfer or augmentation has become a compelling and practical research topic. 

Numerous related works have explored music style transfer to resemble a different music instrument, genre, and artist's style~\cite{dai2018music, cifka2020groove2groove, brunner2018midi, nakamura2019unsupervised, lu2019play, li2024music}. For example, MIDI-VAE~\cite{brunner2018midi} uses a variational autoencoder to perform polyphonic music style transfer, modifying pitch, dynamics, and instrumentation. Another work,~\cite{nakamura2019unsupervised} is introduced to statistically learn melody style transformations without labeled data by clustering pitch and rhythm structures, enabling unsupervised melody conversion. Next, ~\cite{lu2019play} uses timbre representations and GAN-based transformations to enable flexible one-to-many music instrument and genre conversions. Following this, ~\cite{hung2019musical} proposes using disentangled pitch and timbre features to employ instrument transformations while preserving composition structure in a music rearrangement task. Finally, a recent work~\cite{li2024music} introduces a diffusion-based approach that learns example-based stylization by embedding music styles into a pre-trained text encoder, enabling flexible style adaptation from arbitrary reference music audio.

\textit{However, little attention has been paid to the device-aware music transformation}. Existing works~\cite{valimaki2016all, pepe2022deep, martinez2018end, perez2009automatic, borsos2021micaugment, sonowal2022novel, iriz2023coneqnet, hong2018emotional} mainly employ equalization (EQ) methods to boost or reduce specific frequencies (i.e., frequency responses) to achieve a targeted sound balance. Traditional EQ methods, such as parametric EQ, graphic EQ, and shelving filters~\cite{valimaki2016all}, are commonly used to shape audio output, often aiming to match target responses like the Harman curve~\cite{olive2013listener}, which is derived from extensive listening tests to reflect the preferred sound signature of most listeners. Furthermore, underscoring the difficulties of using those handcrafted EQ, several works~\cite{perez2009automatic, martinez2018end, iriz2023coneqnet} leverage simple neural networks to automate equalization, which generally learns a mapping from raw audio to an equalized target or filter parameters. Ultimately, these works focus on conventional music style transfer (e.g., classical to jazz) or EQ, which lacks robustness to unseen devices and do not address \textit{speaker-aware} music transformation. Moreover, they overlook that integrating how the human ear perceives music across various speakers is critical for developing solutions that enhance audio quality or adapt music to the end users. Finally, their modeling approaches lack the perceptual and structured nature of device characteristics in an interpretable way.

In contrast, our work introduces a novel task of \textit{speaker-guided music transfer} using deep learning and also presents a baseline answer to the question: \textit{How to represent device characteristics to achieve effective device-conditioned music transfer?} Specifically, we first collect a music dataset that simulates human ear perception from six earphone and headphone speakers. Subsequently, we propose using a pre-trained vision-language instruction model (VLM) for extracting embeddings of visual frequency response curves (FRCs) from both the speaker and the Harman target~\cite{olive2013listener}. By encoding these FRCs as line graphs, VLM can reason about spectral behavior in a structured and interpretable way, providing rich representations about each device for speaker-aware transformations. These embeddings are fed into our conditional modeling pipeline for device-guided music transfer. Our key contributions are summarized as follows: (1) We present the first work exploring speaker-guided music transfer and introduce a novel music dataset of simulated music playback at the ear level from various headset speakers. (2) We propose \SysName, a framework that uses a vision language model to extract visual embeddings of the frequency responses, conditioning a modified hybrid transformer model. (3) We demonstrate the effectiveness of \SysName through extensive experiments, highlighting its potential for practical device tuning and few-shot adaptation.

\begin{figure*}[t]
    \centering
    \includegraphics[width=0.95\textwidth]{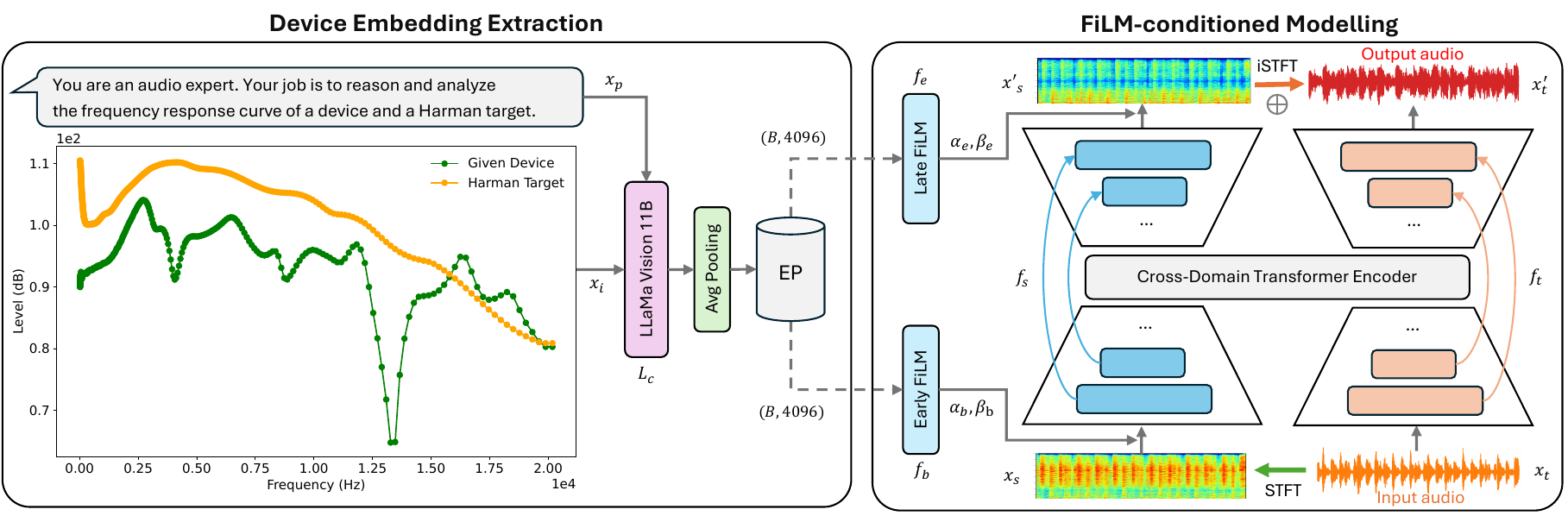} 
    \caption{Illustration of \SysName. Here, EP represents the embedding pool.}
    \label{fig:model}
    \vspace{-0.3cm}
\end{figure*}

\vspace{-0.2cm}
\section{Method}

In this work, we present \SysName, a novel approach for speaker-guided music transfer. As shown in Figure~\ref{fig:model}, our approach leverages frequency response curves (FRCs) as a graphical representation of the devices and VLM-based reasoning to extract structured corresponding embeddings. These embeddings are stored in an embedding pool (EP) and then used to condition a modified hybrid transformer Demucs model~\cite{rouard2023hybrid} via feature-wise linear modulation (FiLM)~\cite{perez2018film}, aiming that the transformed music matches the playback characteristics of a given target device. The subsequent subsections provide a detailed description of key components within our work.

\subsection{Device embedding extractions}

\textbf{Graphical frequency responses}: We first obtain the numerical frequency response (FR) of each device by playing a test signal that spans 480 frequency bands, across 20 to 22050 Hz\footnote{The 480 frequency bands are derived from a common FRC database~\cite{Squiglink} that is used for music production.} and record the output. Subsequently, the response magnitudes are derived by computing the difference in sound pressure level in decibels between the recorded signal and the original input at each frequency band.

Representing FR as a discrete set of abstract numerical values with an unstructured nature makes it difficult for both humans and models to reason about perceptual differences across devices. Therefore, we use it as a line graph format, which aims to provide semantic visual embeddings that encode the device's characteristics. In addition, we include the Harman curve (HC)~\cite{olive2013listener} to provide additional context for more comparable analysis and reasoning of perception perspectives and FR understanding, as the Harman curve is a widely recognized FR and part of the general knowledge of the VLM. 

\textbf{VLM as an embedding extractor}: To extract device embeddings, we propose to use Llama 3-2 11B Vision Instruct, a large-scale VLM trained in a variety of vision and text tasks. It has demonstrated strong reasoning capabilities across structured data, making it a compelling choice for analyzing FR characteristics. Specifically, we provide the VLM model (denoted as $L_c$) with two inputs: an explicit prompt-based instruction ($x_p$) and a line graph image ($x_i$) to guide the reasoning process. Before training, the VLM integrates these inputs and processes a multi-modal representation of the device characteristics. Eventually, we extract the last hidden state of the final decoding layer and then apply an average pooling operation to compress the representation into a fixed 4096-dimensional vector, which serves as the final device representation. 

In particular, to simulate how different users can choose diverse inputs and adapt to the diversity of VLM outputs, multiple embeddings are generated for each device. When training, instead of using a fixed device embedding, we randomly sample one from the available embedding pool (EP) for each device in every iteration, ensuring that the model continuously adapts to slightly different representations of the same device. This design also introduces a flexible manner in the conditioning process, which fixed numerical numbers are inherently unable to achieve.

\subsection{FiLM-conditional modeling}

We propose using FiLM as additional layers to inject those device-specific embeddings into a music-pretrained model. Firstly, we leverage hybrid transformer demucs (Demucs v4)~\cite{rouard2023hybrid}, a special architecture originally designed for music source separation task. It is trained on the MUSDB18 dataset~\cite{rafii2017musdb18}, well-supplemented with an additional 800 curated songs. Specifically, Demucs v4 employs a dual-path design, combining a time-domain and a frequency-domain processing branch (Figure~\ref{fig:model}). The time-domain branch ($f_t$) operates directly on waveforms ($x_t$) using a U-Net encoder-decoder structure with five layers in both the encoder and decoder. The frequency-domain branch ($f_s$) first applies a short-time Fourier transform (STFT) to convert the waveform into a spectrogram, which is then processed through a similar U-shaped structure to extract hierarchical spectral features. The reconstructed spectrogram is transformed back into the waveform via an inverse short-time Fourier transform (iSTFT) and added to the time-domain branch output ($x'_t$). Particularly, a cross-domain Transformer encoder is introduced in the latent space to facilitate interaction between these two representations, with self-attention within each domain and cross-attention across domains. 

To perform device-aware music transformations, we guide the pre-trained Demucs v4 using two FiLM modules ($f_b$ and $f_e$), each consisting of two dense layers that output scale ($\alpha$) and shift ($\beta$) parameters to modulate the model’s spectral features. Here, we employ these parameters only in the frequency branch as the conditioned embeddings encode semantic information about the frequency response. Additionally, we apply FiLM in two positions: the frequency encoder branch (after STFT and normalization layer) and the frequency decoder branch (before iSTFT and denormalization layer). This design benefits the most pre-trained weights of Demucs v4 during tuning while enabling efficient adaptation at a global level, aligning with the fact that VLM-derived embeddings capture broad frequency characteristics rather than fine-grained details for guiding latent or deep feature spaces. 


\section{Data Collection}

\begin{figure*}[ht]
    \centering
    \begin{minipage}{0.3\textwidth}
        \centering\includegraphics[width=0.6\textwidth]{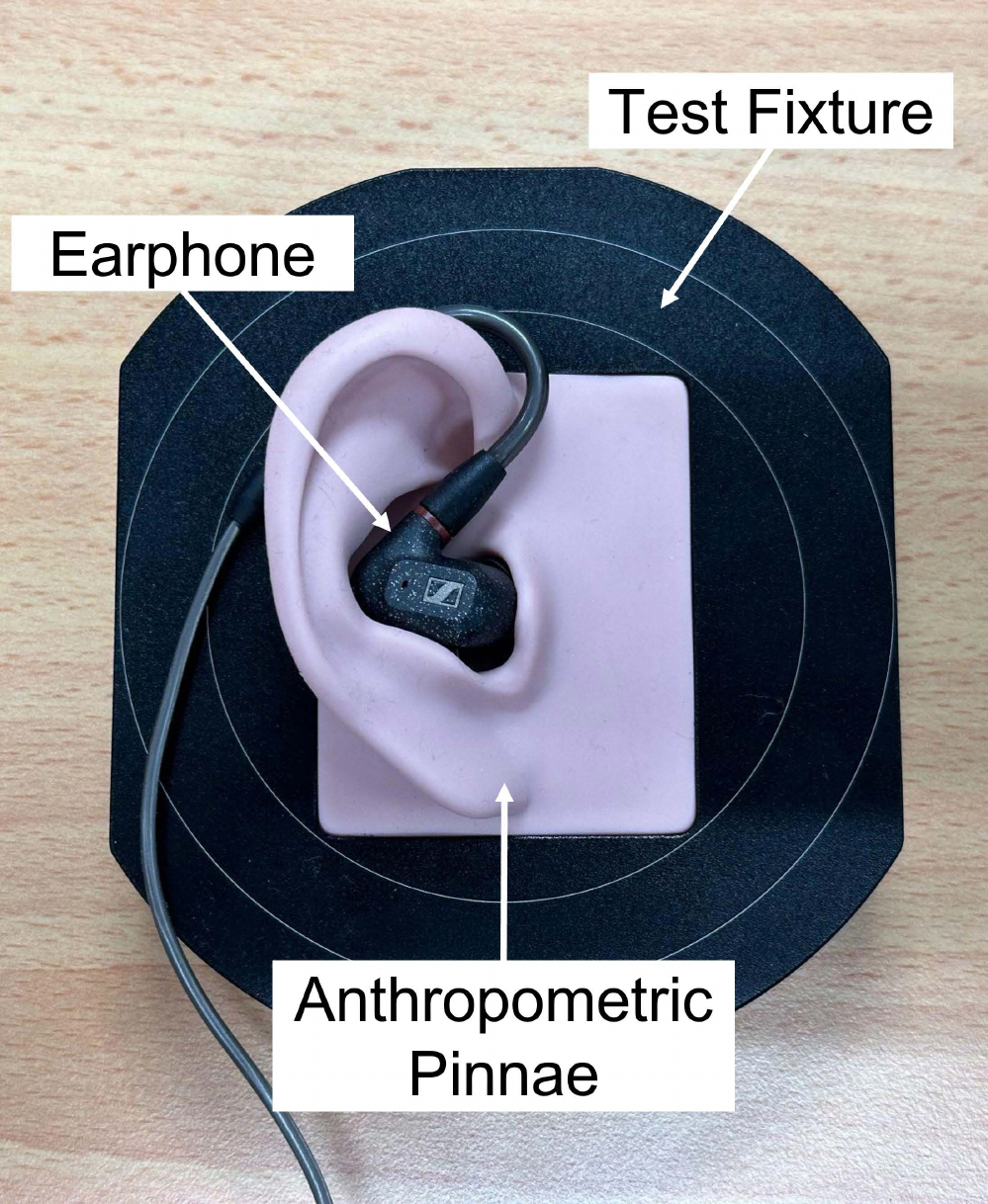}
        \caption{Data collection.}
        \label{fig:setup}
    \end{minipage}
    \begin{minipage}{0.28\textwidth}
    \centering
        \scriptsize
        \setlength{\tabcolsep}{2pt} 
        \begin{tabular}{llr}
            \toprule
            Model                       & Type     & Price   \\ \midrule
            Airline Earphone & In-ear   & Free    \\
            Fransun E20                 & In-ear   & 5 USD   \\
            JBL C100SI                  & In-ear   & 15 USD  \\
            Sennheiser IE 300           & In-ear   & 220 USD \\
            Zihnic Foldable             & Over-ear & 20 USD  \\
            Sony WH-1000XM5             & Over-ear & 300 USD \\ \bottomrule
        \end{tabular}
        \vspace{0.3cm}
        \captionsetup{type=table} 
        \captionof{table}{Earphone and headphone speakers.}
        \label{tab:earphones}
    \end{minipage}
    \hfill
    \begin{minipage}{0.37\textwidth}
    \centering
\includegraphics[width=.97\textwidth]{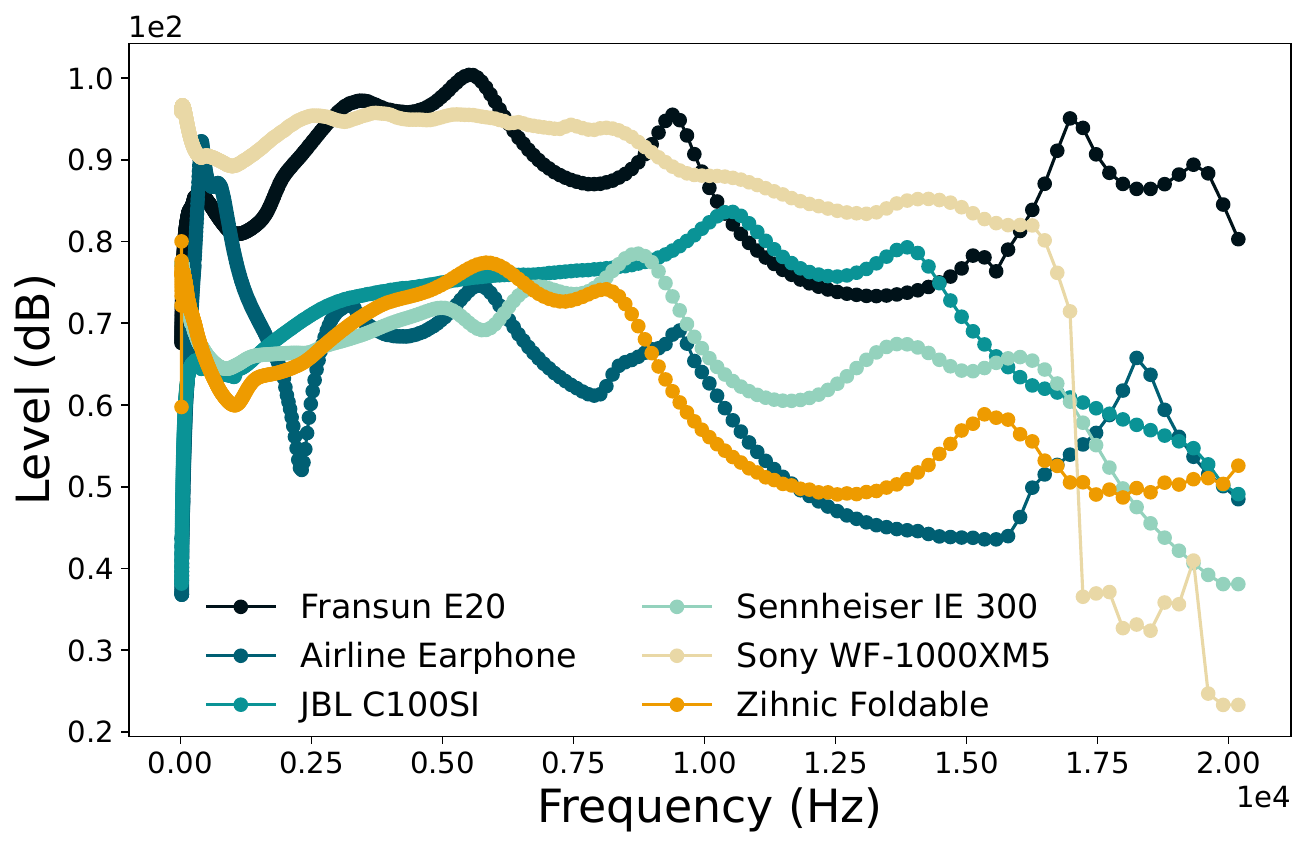}	
    \vspace{-0.2cm}

    \caption{FRCs of the six speakers included in the study.}
    \label{fig:frcs}
    \end{minipage}
\end{figure*}

Training and evaluating \SysName requires a paired music dataset that contains digital input audios and various device-target audios. Here, the targets should capture the playback characteristics of different speakers by simulating real-world listening conditions, particularly at the ear level. However, to the best of our knowledge, there is no existing public music dataset that models these~\textit{in-the-wild} playback systems. To address this gap, we collect a dataset by designing a controlled recording setup that approximates how music is perceived through various devices. 

Specifically, we built a prototype consisting of earphones, an artificial ear model, and an audio development board to \textit{approximate the sound perceived by the human ear}. The artificial ear model (Figure~\ref{fig:setup}) complies with the IEC-711~\cite{international1981occluded} standard, which is widely used to replicate the acoustic impedance and sound transmission characteristics of the human ear. It includes an anthropometric pinna attached to a test fixture, which connects to an ear simulator. A condenser microphone is positioned at the tympanic membrane location within the ear simulator to capture the sound produced during music playback. 

For this prototype, we selected six earphone models covering a wide range of prices, including both in-ear and over-ear types, as shown in Table~\ref{tab:earphones}. The microphone inside the ear simulator and the tested earphones were separately connected to the audio input of a Bela Mini board~\cite{bela} via a 3.5mm audio jack. The Bela Mini board features an integrated development environment (IDE) for audio playback, recording control, and synchronization, and the sampling rate was set to 44.1 kHz.

Following this, we collect the data using the GTZAN Dataset~\cite{tzanetakis2002musical}, which consists of 10 music genres, each containing 100 audio files, each 30 seconds long. For each device, we played and recorded 10 minutes of audio from each genre, resulting in 100 minutes of recorded data per model. With six devices collected, we have 600 minutes of recorded audio in total. It is worth noting that audios are recorded in a close-silent controlled environment to capture the device's unique frequency response characteristics. Examples of the extracted frequency response curves from six devices can be seen in Figure~\ref{fig:frcs}.

\section{Experimental Details}

\subsection{Training configuration}

With the collected dataset, we split each device's data into 60 minutes for training, 15 minutes for validation, and 25 minutes for testing set (no-overlapping files). Subsequently, we apply a windowing process with a segment length of 5 seconds and a stride of 0.5 seconds. Eventually, combining all device data yields 42552, 10638, and 17730 samples for the three sets, respectively. We use an AdamW optimizer with a learning rate of \(5 \times 10^{-5}\) and a weight decay of 0.01 for the model training. The optimizer is configured with \(\beta_1 = 0.9\), \(\beta_2 = 0.98\), and an epsilon value of \(1 \times 10^{-6}\). Training runs for 20 epochs with a batch size of 32. Here, we use the mean squared error (MSE) loss and apply early stopping based on the validation loss. The processes are performed on 4 NVIDIA A100 40GB GPUs.

\subsection{Metrics}

We evaluate the effectiveness of \SysName using three common metrics: Signal-to-Noise Ratio (SNR), Root Mean Square Error (RMSE), and Short-Time Objective Intelligibility (STOI). Here, SNR (dB) measures the ratio of target signal energy to noise, where the higher the better. Next, RMSE measures more the direct error in the time domain, with lower values indicating a closer match to the reference signal. Here, we multiply RMSE with 100 for better interpretation. Finally, STOI (\%), ranging between 0 and 100, originally designed for speech intelligibility assessment yet still provides additional insights into our custom dataset as it indicates correlations in time-frequency representations of the references and predictions.

\subsection{Evaluation configuration}

Using the above metrics, we assess the proposed model and then compare it with the cases when not using VLM-FiLM (no condition) or EP (just single embedding). Here, we also explore the optimal number of stored training embeddings per device in EP (called EP-number) with our proposed setting (EP-30). Note that there is no overlap between these training and test embeddings (fixed 20 for testing). Finally, we investigate the few-shot adaptation capability of \SysName to assess the feasibility of rapid adaptation with minimal training resources. In this experiment, the model is first trained on five devices, leaving out one device as an unseen target. Subsequently, we tune the trained model only on 2\%, and 5\% of the training data while testing the same set from the previous setting.
\begin{table*}[t]
    \tiny
    \setlength{\tabcolsep}{1pt} 

    \centering
    \caption{Evaluation results (SNR in dB, RMSE$\times$100, STOI in \%) across six earphone devices.}
    \label{tab:results}
    \begin{tabular}{c ccc ccc ccc ccc ccc ccc}
        \toprule
        & \multicolumn{3}{c}{Fransun E20} & \multicolumn{3}{c}{Airline Earphone} & \multicolumn{3}{c}{JBL C100SI} & \multicolumn{3}{c}{Sennheiser IE 300} & \multicolumn{3}{c}{Sony WF-1000XM5} & \multicolumn{3}{c}{Zihnic Foldable} \\
        \cmidrule(lr){2-4} \cmidrule(lr){5-7} \cmidrule(lr){8-10} \cmidrule(lr){11-13} \cmidrule(lr){14-16} \cmidrule(lr){17-19}
        & SNR$\uparrow$ & RMSE$\downarrow$ & STOI$\uparrow$ & SNR$\uparrow$ & RMSE$\downarrow$ & STOI$\uparrow$ & SNR$\uparrow$ & RMSE$\downarrow$ & STOI$\uparrow$ & SNR$\uparrow$ & RMSE$\downarrow$ & STOI$\uparrow$ & SNR$\uparrow$ & RMSE$\downarrow$ & STOI$\uparrow$ & SNR$\uparrow$ & RMSE$\downarrow$ & STOI$\uparrow$ \\
        \midrule
        \textbf{DeMT (EP-30)} & \textbf{21.35} & \textbf{0.56} & \textbf{99.51} & \textbf{19.93} & \textbf{0.38} & \textbf{97.66} & \textbf{16.51} & \textbf{0.24} & \textbf{98.88} & \textbf{18.94} & \textbf{0.56} & \textbf{99.43} & \textbf{16.47} & \textbf{0.61} & \textbf{98.66} & \textbf{24.18} & \textbf{0.37} & \textbf{99.30} \\
        \midrule
        DeMT (EP-50) & 19.35 & 0.68  & 99.40 & 18.16 & 0.46 & 96.89 & 12.78  & 0.38  & 98.27  & 16.76& 0.76 & 99.28 & 13.91 & 0.95 & 98.06  & 21.09 & 0.63 & 99.19  \\
        DeMT (EP-40) & 18.25 & 0.80 & 99.40 & 17.92 & 0.48 & 96.80 & 13.22 & 0.38  & 98.73  & 15.82  & 0.81 &  99.02 & 15.09  & 0.73  & 98.62 & 21.71 & 0.51  & 99.31  \\

        DeMT (EP-20) & 17.73 & 0.89 & 99.37 &  15.57 & 0.66  & 97.11  & 11.57 & 0.48 & 98.29 & 16.18 & 0.81 & 99.29  & 14.74 & 0.77 & 98.36 & 19.80 & 0.64  & 99.01 \\
        DeMT (EP-10) & 13.03 & 1.51 & 98.25  & 7.46 & 1.89 & 92.86  & 7.72 & 0.68 & 95.11 & 12.51 & 1.24 & 98.45 & 11.55 & 1.12  & 96.30  & 14.61 & 1.28 & 97.87 \\
 
        DeMT (EP-1) & 6.15 & 3.29 & 96.71 & -1.55 & 4.69 & 89.75 & -3.82 & 2.65 & 94.43 & 9.85 & 1.74 & 97.33 & 6.67 & 2.09 & 96.65  & 3.67 & 3.92 & 93.99 \\
        \midrule

        W/o VLM-FiLM & 3.91 & 4.30 & 98.92  & -0.16  & 3.88  & 95.34  & -4.23 & 2.64 & 98.59 & 6.46 & 2.47  & 98.78 & 4.03 & 2.82 & 97.56 & 3.87  & 3.77  & 97.90 \\
        W/o HC & 19.54 & 0.67 & 99.34 & 18.27 & 0.45 & 96.44 & 15.49 & 0.30  & 98.63 & 17.33 & 0.66  & 99.27  & 16.23 & 0.62  & 98.51  & 18.73 & 0.83 & 98.92 \\
        \midrule
        DeMT (2\%) & 15.01  &  1.18 &  99.08& 12.52 & 1.03  & 95.18  & 10.01  & 0.67  & 95.89  & 15.74 &0.86  & 99.04 & 11.21 & 1.40 & 96.60 & 14.77 & 1.22  & 97.88 \\
        DeMT (5\%) & 16.71  & 1.01  & 99.18 & 14.43 & 0.98 & 95.87  & 10.06 & 0.68  & 96.62 & 15.96 & 0.83 & 98.97 & 11.97 & 1.31 & 97.18 & 16.83 & 1.01  &98.38 \\
        \bottomrule

    \end{tabular}

\end{table*}

\section{Evaluation}

To this end, we constructed a novel dataset (covering six speakers) and processed it into 42,552, 10,638, and 17,730 samples (5-second segments at 44.1 kHz) for the training, validation, and test sets, respectively. In the following sections, we present the main results and various ablation studies in our framework. 

\begin{figure*}
    \centering
    \begin{minipage}{0.34\textwidth}
\centering
\includegraphics[width=0.95\textwidth]{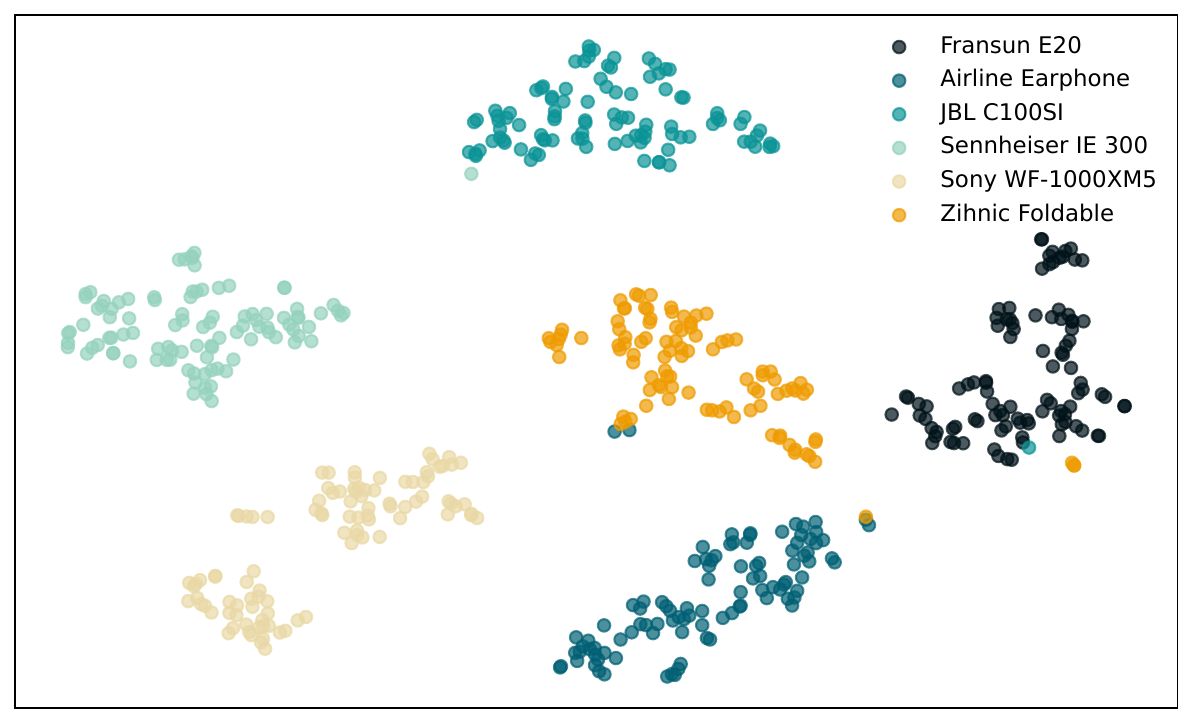} 
    \caption{T-SNE visualization of the device embeddings.} 

    \label{fig:tsne_embeddings}
    \end{minipage}
    \hfill
    \begin{minipage}{0.63\textwidth}
    \centering
    \includegraphics[width=0.95\textwidth]{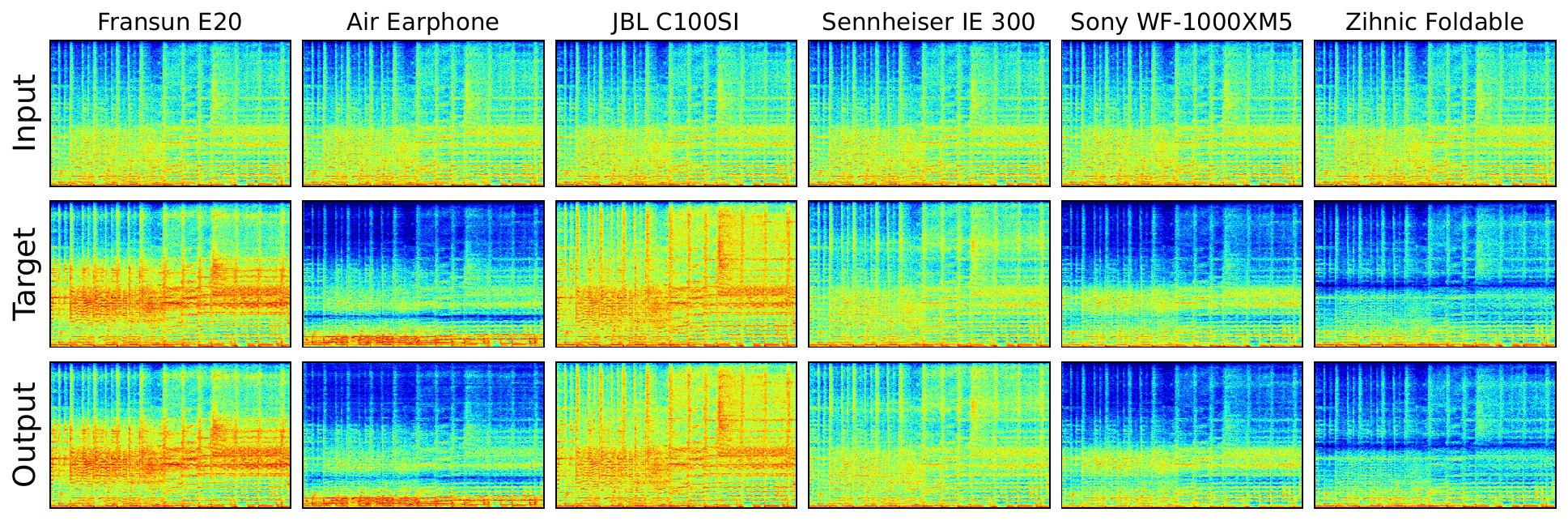} 
    \caption{Examples of input, target, and output audio spectrograms across six audio speakers.}

    \label{fig:specv2}
    \end{minipage}
\end{figure*}

\paragraph*{Main Results.} As shown in Table~\ref{tab:results}, the proposed \SysName (EP-30) achieves strong transfer performance, with an average SNR of 19.56 dB, RMSE of 0.45, and STOI of 98.91\% across six devices. We also show examples of spectrograms in Figure~\ref{fig:specv2}, which illustrates how \SysName effectively reshapes the spectral content to match the characteristics of the target speakers. The points below indicate more results from our ablation studies, including the impact of EP size, VLM-FiLM, and HC.

\paragraph*{Impact of EP size.} First, there is a significant degradation observed when using smaller embedding pool sizes (EP-1, EP-10, EP-20), highlighting the necessity of diverse device representations. In particular, EP-1 results in the poorest performance (SNR of 3.50 dB, RMSE of 3.06, and STOI of 94.81\%), directly demonstrating the importance of utilizing multiple embeddings per speaker. However, extending EP size beyond 30 does not yield further gains, likely due to increased embedding variance leading to underfitting in our current dataset. 

We further validate the quality of our extracted embeddings by evaluating their consistency and differentiation between speakers. Specifically, we visualize the embeddings using t-SNE, as shown in Figure~\ref{fig:tsne_embeddings}. It can be seen with well-formed clusters, indicating that the embeddings remain stable within each device and differ between the six devices. Although there are a few outlier samples, the observed clustering still highlights the potential for broader applications in unseen device adaptation (e.g., possible for zero-shot settings if a broader number of devices were embedded).

\paragraph*{Impact of VLM-FiLM.} Second, without VLM-FiLM, results show a drastic degradation, with the average SNR dropping to 2.31 dB, confirming that the conditional branches play a crucial role in the task. 

\paragraph*{Impact of HC.} Third, the results when not using HC also witness a reduced SNR of 17.50 dB on average, indicating that embeddings extracted from the VLM with HC provide more meaningful information. This is likely because the inclusion of HC as an FC reference term helps the VLM interpret or analyze frequency response curves more effectively and reasonably.

\paragraph*{Generalization to Unseen Devices}

Despite minimal fine-tuning, the few-shot adaptation experiment achieves an average SNR of 13.21 dB, RMSE of 1.06, and STOI of 97.27\% for 2\% of tuning data. Increasing to 5\% further improves performance, reaching an average SNR of 14.32 dB, RMSE of 0.97, and STOI of 97.70\%, showing reasonable generalization to an unseen device. This also suggests the potential for extending \SysName towards more efficient zero-shot adaptation in future work.

\vspace{-0.2cm}
\section{Conclusion}
This work presents \SysName, a novel approach for device-aware music transformation that uses visual frequency response embeddings extracted from LlaMa Vision to condition a FiLM-Demucs hybrid transformer model. \SysName models device speakers using a new self-collected dataset, and experimental results demonstrate that \SysName effectively adapts music playback to target devices while also showing potential for unseen device adaptation, paving the way for future research in device-aware music generation.


\medskip
{\small
\bibliographystyle{plain}
\bibliography{refs}
}



\end{document}